\documentstyle[emulateapj]{article}

\input epsf
\begin{document}

\newcommand{\vertsp}{\vphantom{\displaystyle{\dot a \over a}}}
\newcommand{\se}{{(0)}}
\newcommand{\ve}{{(1)}}
\newcommand{\te}{{(2)}}
\newcommand{\nnu}{\nu}
\newcommand{\Spy}[3]{\, {}_{#1}^{\vphantom{#3}} Y_{#2}^{#3}}
\newcommand{\Gm}[3]{\, {}_{#1}^{\vphantom{#3}} G_{#2}^{#3}}
\newcommand{\Spin}[4]{\, {}_{#2}^{\vphantom{#4}} {#1}_{#3}^{#4}}
\newcommand{\scpot}{{\cal V}}
\newcommand{\tl}{\tilde}
\newcommand{\bm}{\boldmath}
\newcommand{\MNRAS}{Mon. Not. Roy. Astron. Soc.}
\def\bi#1{\hbox{\boldmath{$#1$}}}

\renewcommand{\ell}{l}
\renewcommand{\topfraction}{1.0}
\renewcommand{\bottomfraction}{1.0}
\renewcommand{\textfraction}{0.00}
\renewcommand{\dbltopfraction}{1.0}


\title{Lensing Induced Cluster Signatures in Cosmic
Microwave Background}

\author{Uro\v s Seljak}
\affil{Department of Physics, Jadwin Hall, Princeton University,
     Princeton, NJ 08544}

\author{Matias Zaldarriaga}
\affil{Institute for Advanced Studies, School of Natural Sciences,
Princeton, NJ 08540}
\date{October 1998}


\begin{abstract}

We show that clusters of galaxies induce step-like wiggles
on top of the cosmic microwave background (CMB). The direction of the 
wiggle is
parallel to the large scale gradient of CMB allowing
one to isolate the effect from other 
small scale fluctuations. The effect
is sensitive to the deflection angle rather than 
its derivative (shear or magnification) and is thus tracing outer 
parts of the cluster with higher sensitivity
than some other methods. A typical 
amplitude of the effect
is $10\mu K (\sigma_v/1400 kms^{-1})^2$ where $\sigma_v$ is the 
velocity dispersion of the cluster 
and several $\mu K$ signals extend out to a fraction of a degree.
We derive the 
expressions for the temperature profile for several simple parameterized 
cluster models and identify some degeneracies between parameters.
Finally, we 
discuss how to separate this signal from other imprints on CMB using 
custom designed filters.
Detection of this effect is within reach of the next generation of small scale 
CMB telescopes and could provide information about the cluster density 
profile beyond the virial radius.
\end{abstract}


\section{Introduction}

Fluctuations in the Cosmic Microwave Background (CMB) are
believed to originate from the era of hydrogen recombination 
at a redshift of $z\sim 1100$. Before recombination 
photons and electrons were tightly coupled via the process of 
Thompson scattering, while afterwards electrons were bound to 
protons in hydrogen and photons were allowed to propagate freely 
through the universe. Already before and specially 
during recombination the coupling was not perfect, leading to
erasure of fluctuations in the CMB on small scales. As a result these primary 
fluctuations are expected to be very smooth on scales below 10'.

On very small scales the CMB can be considered as a simple gradient.
A mass concentration in front of such a 
gradient gravitationally deflects the light. 
This deflection causes a fluctuation in the CMB temperature,
which is determined by the mapping between unperturbed and 
perturbed photon position (see also Kosowsky et al. 1999).
This small scale power is preferentially 
generated in the regions of high gradient of primary CMB anisotropies.
The effect can be generated by any mass concentrations
along the line of sight, 
such as galaxy halos, clusters and superclusters. In this paper we 
concentrate on clusters, which being massive and big may generate
a particularly strong effect.
They are thus the primary
candidates for detection of this effect on individual objects, as 
opposed to the statistical detection discussed in 
Zaldarriaga and Seljak (1999a,b).
The purpose of this paper is to analyze their imprint on the microwave sky
by analyzing a number of simple cluster profiles and
discuss its detectability for realistic observational scenarios.

It should be stressed that 
this gravitational lensing effect is different from the lensing effect of 
a cluster
discussed in Zaldarriaga and Seljak (1999a). 
There CMB was viewed as a collection of peaks,
with a well determined distribution of shapes and sizes in Gaussian 
models. These will be 
distorted as they pass by a large massive object, generating 
a coherent ellipticity or size distortion which can be identified
by averaging over sufficient number of independent patches. 
By averaging over the CMB the lensing effect 
can be isolated and a cluster density profile can be reconstructed 
Zaldarriaga and Seljak (1999a). In practice this requires the presence of small 
scale CMB fluctuations at detectable levels and these are not likely 
to originate from primary anisotropies. Secondary processes and foregrounds
reviewed in this paper could provide the small scale power required, although 
the level of these small scale fluctuations is still uncertain at the
present. 
In principle this would provide an alternative method to reconstruct the 
cluster density profile in addition to the one discussed here. Given the 
uncertain level of secondary anisotropies we will in the remainder of 
this paper ignore this possibility, adopting a conservative 
position that secondary anisotropies 
are only a source of confusion to the signal one 
is trying to isolate. 

\section{Lensing effect of cluster on CMB}

The measured temperature field $T({\bi \theta})$ at observed position 
$\bi{\theta}$ originates 
from some unlensed position ${\bi \theta}'$ of
the CMB field at the last scattering surface
$\tl T({\bi \theta}')$. The relation between the two 
is given through the deflection angle of the CMB photons
$\delta {\bi \theta}$,
\begin{eqnarray}\label{expansion}
T({\bi \theta})&=&\tl T({\bi \theta}')=
\tl T({\bi \theta}-\delta {\bi \theta})
\nonumber \\ &\approx& \tl T ({\bi \theta}) -\delta {\bi
\theta}\cdot \nabla \tl T({\bi \theta}).
\end{eqnarray}
In the second line we expanded the temperature using
a linear expansion valid on scale below the coherence 
length of the CMB gradient, which is of the order of 15' for typical models
in a flat universe. On scales below that, 
the primary anisotropies are expected to have negligible power. In this 
case we can treat the unlensed temperature field as a pure gradient.
We are ignoring
all the secondary anisotropies and foregrounds generated along the line of 
sight that will contribute to fluctuations on these small scales.
These act as a source of noise and are discussed later in the paper.

We choose
${\bi \theta}=(\theta_x,\theta_y)=
(\theta \cos \phi_{\theta},\theta \sin \phi_{\theta})$ to be
the observed
position in the sky with origin at the cluster center. The derivative 
of deflection angle with respect to $\bi{\theta}$ is the shear tensor, which 
can be decomposed into its trace part, $2\kappa$,
and two shear components $\gamma_1$ and $\gamma_2$. 
The convergence $\kappa$ is dimensionless and can be expressed
in terms of projected density $\Sigma$ as $\kappa=\Sigma/\Sigma_{cr}$, where 
$$\Sigma_{cr}={c^2 D_{OS} \over 4\pi G D_{OL}D_{LS}}.$$
Here $D_{LS}$ is the angular diameter 
distance from the lens to the source, $D_{OS}$
that between the observer and the source and $D_{OL}$ between 
observer and lens. 
We may parameterize the density profile of the cluster 
in units of a characteristic length scale $r_s$ as 
$\rho(x)$, where where $x=r/r_s$ and $r$ is the radius.
When we measure angles in units of $\theta_s=r_s/D_{Ol}$, so that 
$x=r/r_s=\theta/\theta_s$, 
the deflection angle scales as $\delta \theta \propto m(x)/x$, where 
$m(x)$ is the mass enclosed within the projected radius $x$. 

\begin{figure*}
\begin{center}
\leavevmode \epsfxsize=6.0in \epsfbox{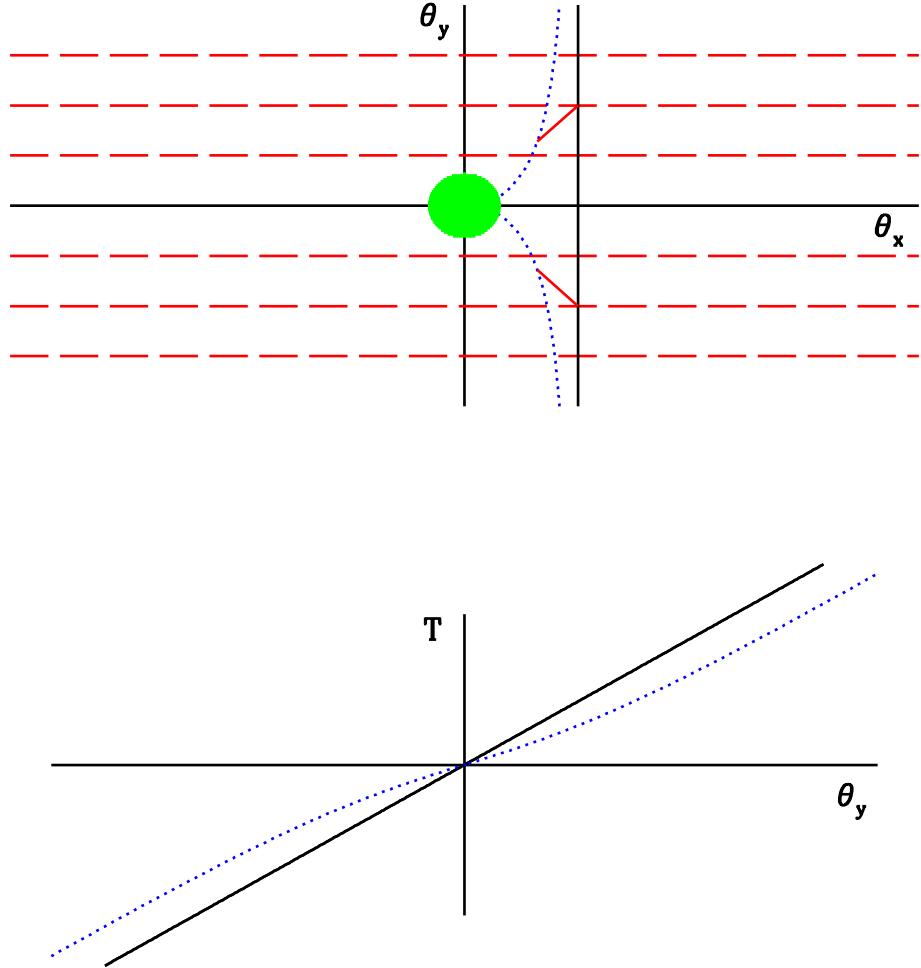}
\end{center}
\caption{In the upper panel we show a cluster lensing a background
gradient. The displacement is radial and a photon observed along the 
fixed $\theta_x$ direction (solid line) is originating from a 
different position behind the cluster (dotted line),with a different 
value of the CMB anisotropy. The bottom panel shows the temperature measured for a
fixed $\theta_x$ as a function of $\theta_y$ in the presence and
absence of the cluster. Points with $\theta_y>0$ get deflected to
a smaller $\theta_y$ in the lens plane and thus for a positive
gradient they will have a lower temperature when the cluster is
present.
The opposite is true if $\theta_y<0$.}
\label{sketch}
\end{figure*}

Without a loss of generality we may take the gradient to be along
the $y$ axis with an amplitude $\tl T_{y0}$. The observed
temperature becomes,
\begin{equation}\label{tunlensedcl}
  T(\bi \theta)=\tl T_{y0} (\theta_y - \delta \theta_y).
\end{equation}
In the absence of deflection $\delta \theta_y$ one would 
measure  a pure gradient. Any small scale deviation from it is a 
signature of the deflection $\delta \theta_y$. If we measure the value
of the large scale gradient by filtering out 
small scales contaminated by the cluster lensing 
we would know where 
a certain value of the CMB anisotropy
should have come from in the absence of deflection. 
The difference between the expected and measured position is a 
direct measurement of $\delta \theta_y$ and so of the gravitational 
effect of the cluster. 
The effect of lensing by a cluster on the CMB 
can be understood with the help of figure
\ref{sketch}. In the absence of lensing we
would observe just the gradient. Because of the lensing effect by the cluster, 
the light rays will be deflected radially 
so that for $\theta_y > 0$ the rays are
coming from a lower value of $\theta_y$ at the last scattering
surface. 
If the gradient is positive this implies that for
$\theta_y > 0$ in the presence of the cluster we would observe a
lower temperature than what would be observed if the cluster was
not there. The opposite is true for $\theta_y < 0$. Far away from
the cluster the lensed temperature should coincide again with the
gradient. Thus the cluster creates a wiggle on top of the large
scale gradient.

It is important to stress that
the method proposed here is sensitive to one component of the
deflection angle and not the shear or magnification as is the case
for the usual weak lensing reconstruction from background galaxy 
ellipticities or magnitudes. 
It is sometimes argued that we can never measure
the deflection angle in a lensing system because we do not know
the original position of the background image. In this case we can
get around this argument because we know that the background image
is a gradient which we can measure on scales larger than
the cluster. Although both shear and deflection angle are sensitive
to the cluster mass profile the latter involves one derivative of
gravitational potential less than the former. As such it is
less sensitive to small scale fluctuations in the cluster profile 
and more sensitive to outer parts of the cluster, as discussed below.

Another important point is that the effect discussed here is
proportional to the gradient $T_{y0}$. This provides a unique
signature which we
may use to separate it from
other sources of anisotropies.
It also implies that one should select the clusters on which to 
look for this effect not only on the basis of the strength of the
gravitational lensing signal, but also on the basis of the amplitude
of CMB gradient at that position. MAP or some other 
CMB experiment with 15' resolution could provide such information.

\subsection{Singular isothermal sphere}

For a singular isothermal sphere the 
density scales as $\rho \propto r^{-2}$. In this case the
deflection angle is constant,
\begin{equation}\label{defcluster}
  \delta \theta_y=b {\theta_y \over \theta},
\end{equation}
where $b=4\pi (\sigma_v/c)^2
D_{LS}/D_{OS} \sim 1'(
\sigma_v/1400km/s)^2D_{LS}/D_{OS}$, 
where $\sigma_v$ the cluster velocity dispersion.
With the source 
at $z \sim 1100$ we may assume $D_{OS} \gg D_{LS}$, which 
in an Einstein De-Sitter
universe gives $D_{LS}/D_{OS}\approx 1/\sqrt{1+z_L}$.

\begin{figure*}
\begin{center}
\leavevmode \epsfxsize=6.0in \epsfbox{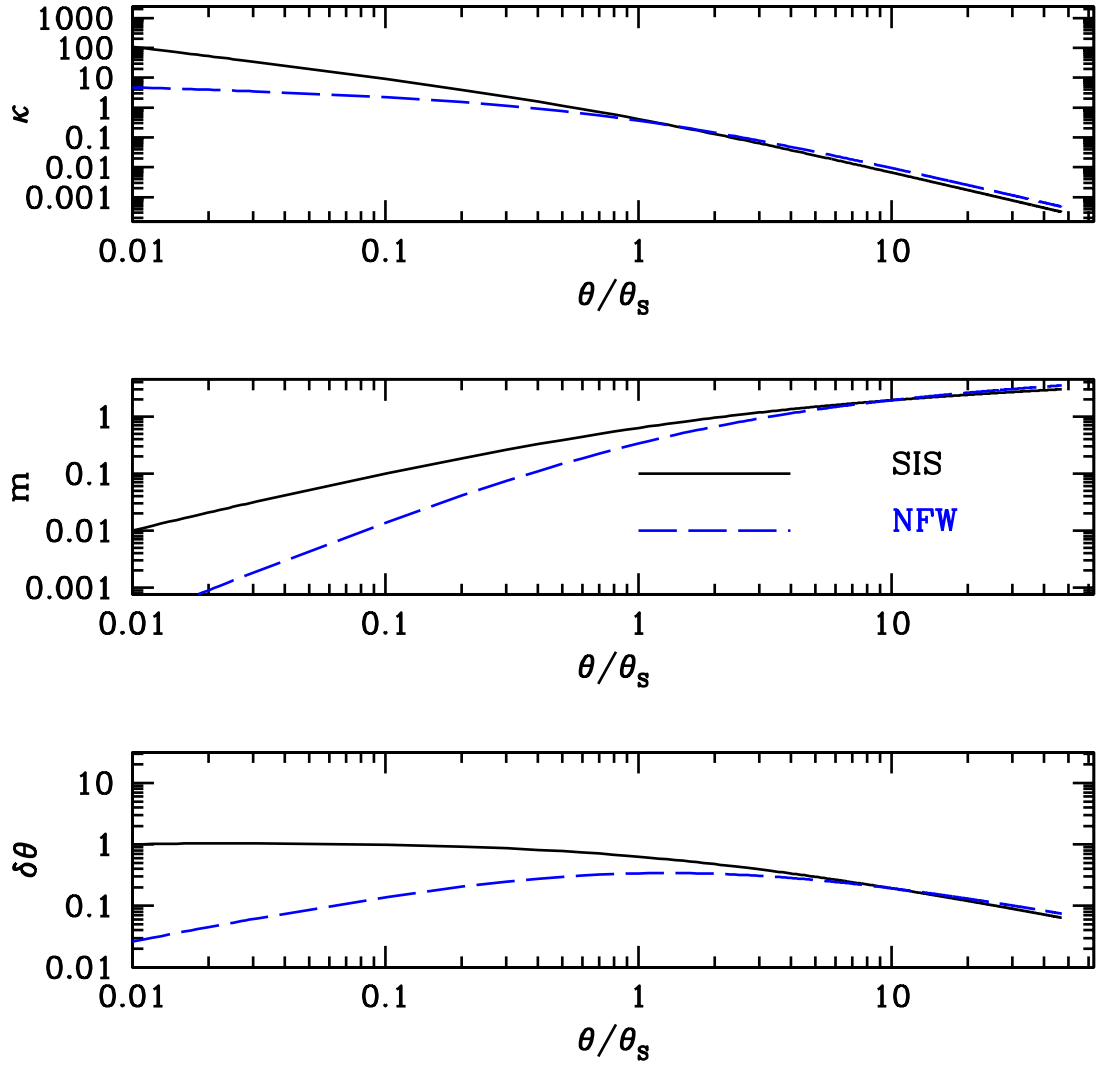}
\end{center}
\caption{Dimesionless surface density $\kappa$, mass $m$ within the 
projected distance and deflection angle $\delta \theta$ as a 
function of projected distance $x=\theta/\theta_s$. Solid lines are for
SIS and dashed for NFW profile.
}\label{fig2}
\end{figure*}

The mass for a singular
isothermal profile grows linearly with radius so at outer parts
the profile must turn over to a steeper slope.  
We will adopt the profile
\begin{equation}
\rho(x) \propto [x^2(1+x)]^{-1},
\end{equation}
where the slope in the outer parts of the cluster has been matched to the 
NFW profile discussed below.
We have numerically integrated 
the equations above to compute the mass within a given radius and 
the deflection angle. These are shown together with the surface 
density $\kappa$ in figure \ref{fig2}, where it can be seen that $\kappa$
is dropping with radius much more rapidly than $\delta \theta$ is. 

\begin{figure*}
\begin{center}
\leavevmode \epsfxsize=6.0in \epsfbox{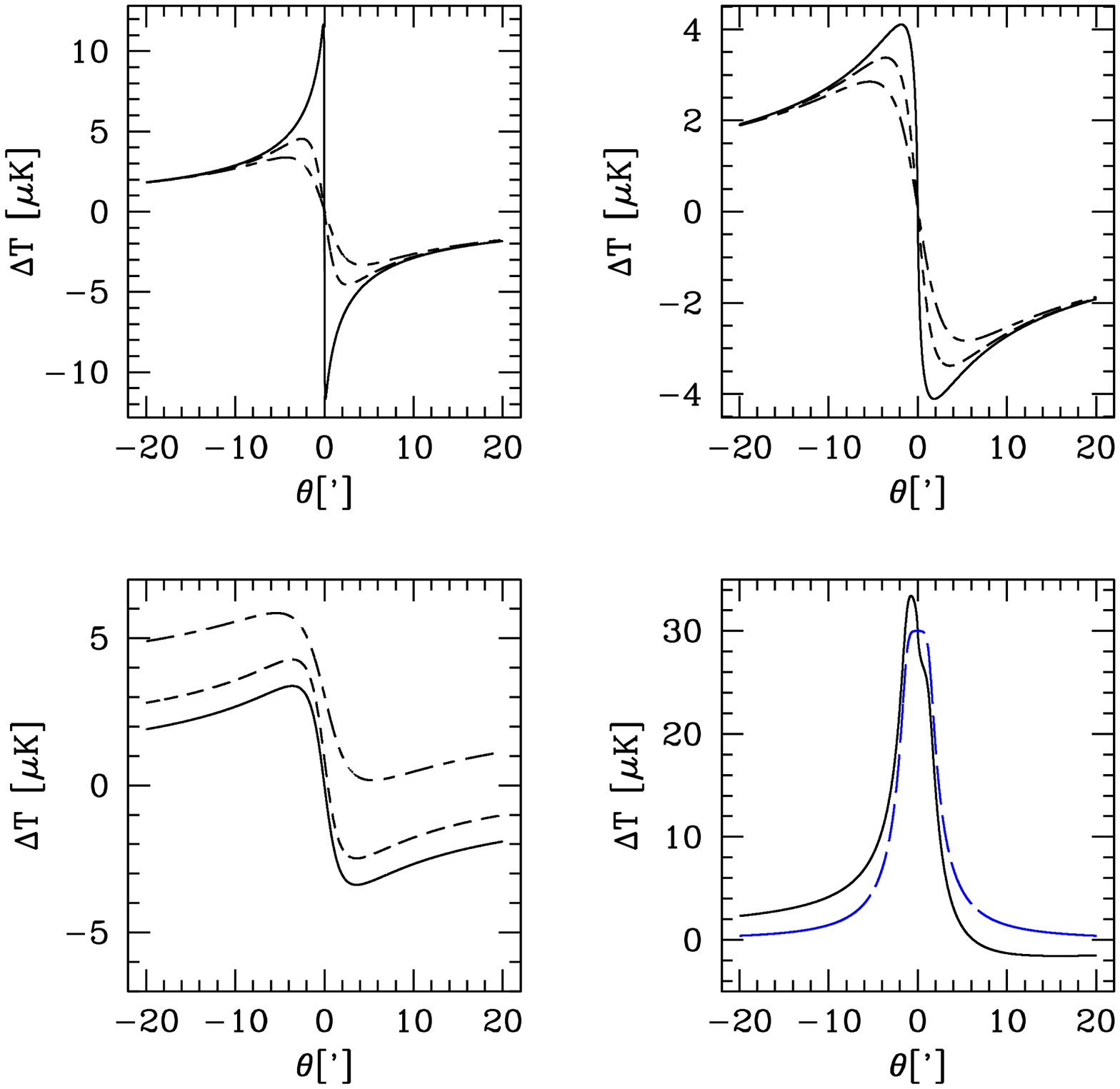}
\end{center}
\caption{Temperature profile of a CMB gradient lensed by a
SIS (upper left) and NFW (upper right) cluster at a fixed $\theta_x$. 
From top to bottom are $\theta_x=0,3,10'$.
The gradient part has been subtracted out for clarity. 
Bottom left shows the same profile for NFW with external shear, 
showing how the odd symmetry across the $y$-axis is broken for $\theta_x \ne 0$.  
Bottom right shows imprint of kinetic SZ (dashed) and 
lensing effect and kinetic SZ superimposed (solid). This also 
breaks the symmetry across the $y$-axis, but it is more concentrated in the 
center than the effect of external shear. 
}\label{fig1}
\end{figure*}

In figure \ref{fig1} we show the signature of the effect on the 
CMB itself. We have subtracted out the gradient term.
We focus on the temperature as a function of
$\theta_y$ for a fixed $\theta_x$. 
We adopted  
$\sigma_v=1400km/s$ and 
$\theta_s=r_s/D_{OL} \sim 1.4'$ (corresponding to A370, see Williams, Navarro and 
Bartelmann 1999).
The amplitude of the wiggle is
$T_{y0}\delta\theta$, proportional to the amplitude of the
gradient and the deflection angle. 
For $\theta_x=0$ the distortion caused by SIS cluster 
would be constant and negative 
for $\theta_y>0$ and constant and positive for $\theta_y<0$, with 
a step function at $\theta_y=0$, reflecting the absence of a core in 
this model. The change in slope in outer parts alters this prediction,
so that only for $\theta<\theta_{s}$ is the deflection angle approximately
constant (figure \ref{fig2}).
For $\theta_x \ne 0$ the temperature profile is smooth, 
but the functional dependence still has odd symmetry with 
respect to the transformation across the $y-$axis, as shown in 
figure \ref{fig1}a.  
The value for the distortion depends on the amplitude of the 
large scale gradient which has an
rms value $\sigma_{\nabla T} =<T_x^2+T_y^2>^{1/2}$ 
of the order 13$\mu K{\rm arcmin}^{-1}$ for standard CDM.  
Other models that fit the current observations give similar 
values for $\sigma_{\nabla T}$. We have adopted this value of the
gradient for our calculation which
gives a 
distortion $\Delta T
\sim b \sigma_{\nabla T} \sim 13\mu K (\sigma_v/1400km/s)^2D_{LS}/D_{OS}$. 
Note that $\mu K$ signals can be 
obtained well beyond the virial radius and that by averaging over the 
entire profile of the signal, one can significantly reduce the 
level of contamination from other contributions.
This is discussed
in more detail in \S 4, where we address more generally
the observability of this signal.

\subsection{NFW profile}

Navarro, Frenk and White (NFW; 1997) proposed a 
universal mass profile that was shown to
fit most of the halos in cosmological N-body simulations.
Its 3-d form is particularly simple and is given
by 
\begin{equation}
\rho(x) = {\rho_s \over x(1+x)^2},
\end{equation}
The transition between $r^{-1}$ scaling in the center and $r^{-3}$ outside
is governed by the scale radius $r_s$. Typical numbers for a cluster halo 
are of the order of $250h^{-1}$kpc for $r_s$, which is
about 15-20\% of the virial radius $r_{200}$, 
defined as the radius where
the mean overdensity inside it is 200.

We numerically calculated the expected temperature profile for the 
parameters of A370, shown in figure \ref{fig1}b. We normalized both
SIS and NFW profiles to have the same total mass at large radii. 
Because of our choice of normalization, 
we see that the largest difference between the NFW and SIS profiles occurs
near the core of the cluster, in the inner $2'$. The NFW profile 
has much less mass near the center and thus has a much smaller
deflection angle. Arcminute or better resolution is needed to
distinguish these two profiles with this method. We will show below 
that other contributions such as instrument noise, 
infrared (IR) or radio point sources
as well as SZ emission further complicate this separation.

\subsection{Non-axisymmetric profile}

Let us now introduce a 
quadrupole deviation from axial symmetry in the form of 
external shear. This can be parameterized with 2 components,
distortion along the $x,y$-axis parameterized with 
$\gamma_1$ and distortion along the diagonals parameterized with $\gamma_2$.
Fermat's gravitational potential can be parameterized
in the following form
\begin{equation}
\Phi={{\bi \delta \theta}\cdot {\bi \delta \theta} \over 2}
-f(\theta)- {1 \over 2} \theta^2 [\gamma_1 \cos 2\phi_{\theta}
+\gamma_2 \sin 2\phi_{\theta}],
\end{equation}
where $f(\theta)$ is a general function describing the axi-symmetric radial
profile of the projected cluster potential.
This can be expanded into a series $f(\theta)=f_1\theta+
f_2\theta^2/2+...$.

From Fermat's principle we obtain,
\begin{eqnarray}
\nonumber \\
\delta \theta_y=\theta_y\Big(\gamma_1-{f'(\theta) \over \theta}\Big)-
\gamma_2\theta_x
.
\end{eqnarray}
Inserting the expansion 
of $f$ above  
we find that $\gamma_1$ and $f_2$ are degenerate, since 
$\delta \theta_y$ has the same dependence on $\bi \theta$ for both parameters.
This degeneracy is similar to the mass sheet degeneracy that exists in the case
of cluster reconstruction from the shear. In that case a constant mass sheet 
cannot be detected using the shear information alone. Similarly here we cannot
separate between a constant mass sheet and an external shear component 
$\gamma_1$. A more general form of this degeneracy is derived in the next 
section. 

External shear   
distortion that is not 
perpendicular or parallel to the $y$ axis can be measured from the 
profile.
The case of $\gamma_2 = 0.3$ and an NFW profile is shown 
in figure \ref{fig1}c. 
Non-axial symmetry breaks the odd parity symmetry across y-axis for 
$\theta_x \ne 0$. At a given $\theta_x$ the whole wiggle is moved up or 
down depending on $\gamma_2$ (of course far away from the shear source 
it is restored back to the unperturbed value). 
This is not the only effect that can break this symmetry.
As discussed in more detail in \S 4 kinetic and 
thermal SZ effects also imprint a signal in the CMB. Kinetic
SZ in particular cannot be distinguished from lensing or primary 
CMB on the basis of frequency dependence. For an axial-symmetric 
cluster it produces a profile with even parity across y axis.
This effect combined with the lensing effect also breaks the 
symmetry as shown in figure \ref{fig1}d. It is much more
centrally concentrated than the effect of external shear, so that the 
two can be separated.

\section{Reconstruction of projected density}

Rather than parameterizing the surface density of the cluster
one may also attempt to reconstruct it directly. To do this we first subtract 
from the CMB anisotropies the pure gradient term and divide 
the temperature by $T_{y0}$. This gives an estimate of $\delta \theta_y(\bi{\theta})$.
We then Fourier transform it 
$\delta \theta_y(\bi{l})=\int d^2l e^{i\bi{l}
\cdot \bi{\theta}}\delta \theta_y(\bi{\theta})$.
The dimensionless surface density $\kappa(\bi{\theta})$ is given by 
inverse Fourier transform of 
\begin{equation}
\kappa(\bi{l})= {-il^2\delta \theta_y(\bi{l}) \over l_y}.
\end{equation}
This inversion is possible for all modes except $l_y=0$. These are long
wavelength modes in the $y$ direction that cannot be distinguished 
from the CMB gradient itself. Hence the inversion
is not unique, although the number of modes for which the inversion fails is 
small compared to their total number (and becomes a set of measure 0 in the 
limit of perfect resolution). This degeneracy is similar to the mass-sheet
degeneracy, which prevents one from reconstructing $\kappa$ from
ellipticity data 
for $\bi{l}=0$ mode. It is more severe here, because there is a whole vector 
of modes for which the inversion fails, rather than just a single mode.
In the previous section we discussed a particular example of this degeneracy 
which prevents one from distinguishing external shear in the direction 
parallel or 
perpendicular to the CMB gradient from a constant surface density term.
Furthermore, even if $l_y \ne 0$,
for modes with small $l_y$ and large $l_x$ this reconstruction 
amplifies any noise contribution present in the data, so the final
map no longer has uniform noise properties.

\section{Sources of noise}

To analyze whether the
theoretical predictions above can be detected we need
to compare them to various sources of noise. 
These sources can 
be divided into instrumental and astrophysical. 
Astrophysical sources can arise from earth atmosphere,
our galaxy, various cosmological sources along the line of sight
and from the cluster itself. They can also arise from gravitational 
lensing by other objects along the line of sight. 
Another source of noise is the CMB
itself, in the form of its deviation from a pure gradient form
assumed in the reconstruction. We may parameterize these sources
of noise with their power spectrum, which will characterize the 
level of fluctuations as a function of scale. This does not 
contain all the information for a non Gaussian process, 
but most of the noise sources that are accumulated along the line of 
sight will be well approximated as Gaussian because of the projection.
Others, such as primary CMB, are believed to be Gaussian already. 
The most important source of noise which cannot be described with
power spectrum information
is emission from the cluster itself, specially 
SZ and dust. We will 
discuss these sources of noise in more detail below.

\subsection{Signal to noise analysis}
The total CMB anisotropy can be modeled as
\begin{equation}
\Delta T (\bi{\theta})=Ag(\bi{\theta})+n(\bi{\theta})
\end{equation}
where $g(\bi{\theta})$ is the angular profile of the deflection 
angle normalized to unity at $\theta=\theta_s$.  
For axi-symmetric clusters its form can be simplified to 
$g({\theta})\cos(\phi_{\theta})$,
where $\phi_{\theta}$ is the azimuthal angle of $\bi{\theta}$.
$A$ is a constant that includes both the strength of the cluster and 
the magnitude of CMB gradient, while the noise term 
$n(\bi{\theta})$ denotes all 
the other contributions to the measurement. 
They can be parameterized with 
the power spectrum 
\begin{equation}
\langle n(\bi{l})n(\bi{l}')\rangle=C(l)\delta(\bi{l}+\bi{l}').
\end{equation}
If we have some knowledge on the 
profile of the cluster deflection angle $g(\bi{\theta})$
we can average the temperature over this profile, thus 
reducing the noise contribution from other sources that 
do not correlate with the expected profile. 
We wish to  
derive the filtering function $\Psi(\bi{
\theta})$ with which we process the data to obtain an estimate of
$A$,
\begin{equation}
\hat{A}=\int \Psi(\bi{\theta})\Delta T(\bi{\theta})d^2\theta.
\end{equation}
We can vary the filtering function with respect to signal to 
noise estimate $A/\sigma$, where 
\begin{equation}
\sigma^2=\langle (A-\hat{A})^2\rangle=\int |\Psi(\bi{l})|^2C(l)d^2l.
\end{equation}
It can be easily shown that the optimal filter is 
$\Psi(\bi{l}) \propto g(\bi{l})/C(l)$ and that
the variance for this filter is given by (Haehnelt \& Tegmark 1996)
\begin{equation}
\sigma=\Big[\int {|g(\bi{l})|^2 \over C(l) }d^2l
\Big]^{-1/2}.
\end{equation}
If the noise spectrum is white then the profile of 
the filter is simply the profile of the deflection angle. This is 
what one expects, since in that case one is obtaining a positive 
value. If however
the noise power spectrum has more power on large (small) scales then 
those large (small) modes are more important to suppress than the small (large) 
scale modes. To suppress large scale modes one has to design a filter that 
is oscillating, so that its shape cancels the slowly changing mode, while still
maximizing the information from the cluster profile. This 
is what is achieved with the optimal filter above.

For an axi-symmetric cluster the Fourier transform simplifies to
\begin{equation}
g(\bi{l})=g(l)\cos \phi_l \,\,\,\,\,\,\,\,\,\,\,\,\,
g(l)=2\pi \int \theta g(\theta)J_1(l\theta)d\theta
\end{equation}
where $J_1(x)$ is the Bessel function of first order and $\phi_l$ is the 
azimuthal angle of $\bi{l}$. The variance becomes
\begin{equation}
\sigma=\Big[\pi \int {g(l)^2 \over C(l)}ldl\Big]^{-1/2}.
\label{sigma}
\end{equation}

In the examples below we will use NFW profile for the cluster, observed
out to a given radius $\theta_0$. We will use $\theta_s=1.4'$ and 
use various sources of confusion to estimate $\sigma$. This can be 
compared to the expected $A$ for large clusters, of the 
order of 5-10$\mu K$, to identify 
the main sources of noise and the range over which this 
method could be used to study the clusters.

\subsection{Instrument degradation}

The detector adds 
noise to the signal. This can be parameterized by its power spectrum, 
which for many instruments can be approximated as a constant,
$C_n(l)=\sigma_n^2 \Omega_p$, where $\sigma_n$ is the rms noise
at each pixel and $\Omega_p$ is its solid angle. Current 
observations of SZ at 30GHz are reaching noise levels of order of 
$\sigma_n=15 \mu K$ at 1-2' resolution (Carlstrom et al. 1999). 
This is approaching the level of the signal predicted here, although 
at these frequencies the dominant signal in the center is coming from the
SZ effect.
Interferometers  are not 
sensitive to low spatial frequency modes, so one cannot obtain the 
direction of the gradient from the experiment itself. This must be 
obtained from a lower resolution experiment such as MAP. 
It is possible that the characteristic signature generated 
by the lensing effect could be observed at larger radii even at 
these low frequencies and 
it would certainly be worthwhile to integrate a few of the clusters 
down to 5$\mu K$ in search of this effect, specially once the direction 
and the amplitude of the gradient of CMB is better known.
At 217 GHz, which is the zero crossing frequency for thermal SZ, 
current observations only 
reach 100 $\mu K$ noise per pixel at a similar resolution (Church et al. 1997),
dominated by atmosphere noise discussed below. 
The next generation of small scale 
experiments with larger arrays and longer observation times 
such as MINT or ACBAR will have the sensitivity
reaching 5 $\mu K$ per arcminute size pixel in a 100 pixel array over a month of 
observation and will be more suitable to detect this effect.

Because of the finite
angular resolution of the instrument 
the predicted $\Delta T/T$ has to be convolved with
the window function of the beam. This dilutes sharp features
around the center of the cluster, such as those produced by SIS
in figure \ref{fig1}a. For this case arcminute resolution would be 
desirable. For less steep profiles such as NFW this is less important. 
Since the actual signal extends quite far away from the center
even a modest resolution of several arcminutes would still be useful,
assuming that other sources of noise discussed below do not dominate 
the signal. 
To incorporate the beam
dilution into the formalism above we may replace $C_n(l)$ with 
$C_n(l)\exp[\theta_b^2l(l+1)]$, where $\theta_b$ is the Gaussian width 
of the beam.
Note that beam dilution does not affect the signal to noise
if noise is not dominated by instrument noise. 

Applying the noise power spectrum to the equation (\ref{sigma}) 
we find that a 10'$\times$10' array with 100 pixels, each with 
$5 \mu K$ noise gives
rms noise $\sigma= 1.6\mu K$ if the effect of the beam is negligible.  
For beam with 1' FWHM this number increases to 3$\mu K$. Doubling the size
of the array or halving the noise per pixel both reduce this number by 
roughly one half. These levels of noise are therefore necessary for
a positive detection of the effect. Note that doubling the size of array
and keeping the noise per solid angle fixed (equivalent to keeping 
the observing time fixed) does not change significantly 
the rms variance. This is 
because the signal is only slowly dropping off with distance from the 
center. In this case changing FWHM from 1' to 2' makes almost no 
difference. For Planck 217GHz channel with 12 $\mu K$ sensitivity 
per pixel
and 5' FWHM the noise level is of the order of 10 $\mu K$, which is 
at the detection limits for large clusters. Except for a few exceptional 
cases Planck will therefore not be able to detect this effect.

\subsection{Intrinsic CMB fluctuations}

We have assumed throughout that the CMB can be approximated as 
a pure gradient. Typical coherence length for CMB gradient is 15' and since
the lensing effect extends well beyond this scale this
approximation breaks down at large separations from the 
cluster center. To estimate the level of this contribution we
can use the CMB power spectrum as a source of noise in equation (\ref{sigma}).
Because the large scale CMB is approximated as a gradient and 
removed we exclude
the modes larger than the size of the observed field. The modes smaller 
than the size of the box cannot be approximated as a gradient
and they contribute noise, which needs to be distinguished from 
the cluster signal. Without removing the long wavelength 
modes with the optimal filter the rms contribution from CMB is 
of the order of 15 $\mu K$ for a survey of $10'\times10'$ increasing
to 100 $\mu K$ for $1^{\circ}\times 1^{\circ}$. 
For such large fields
long wavelength modes of CMB are a significant source of
confusion. 
This reflects the strongly 
correlated nature of CMB on large scales. 
The optimal filter suppresses the influence of 
long wavelength modes by employing alternative 
positive and negative
radial weights. This significantly reduces the
long wavelength modes, while still preserving to a large extent 
the information about the 
cluster profile. In this case
the variance decreases significantly to 4-5$\mu K$. For 
Planck 217 GHz channel the total variance
remains 10 $\mu K$, dominated by the detector noise rather than CMB.
The dominant contribution to the CMB gradient comes from $l > 500$ with
50\% contribution coming from $l>1000$. 
Degree size fields may thus have a significantly lower rms CMB gradient than 
10' size fields. For large areas the best strategy is to select fields with
a smooth and large gradient across the entire field, thus enhancing the signal 
and reducing the level of CMB noise.   

\subsection{SZ effect from the cluster}
SZ effect is the dominant signal from clusters in the low frequency range.
It is caused by scattering of photons by hot electrons in the cluster.
The net effect is to increase the energy of photons and since their number
is conserved this causes their
redistribution from the low frequency Rayleigh-Jeans regime into 
the high frequency Wiener regime. This creates a deficit 
of photons and therefore a CMB decrement at low frequencies and an
increment at high frequencies, with zero crossing at 217 GHz. 
The amplitude of the effect is proportional 
to the temperature of the cluster and its optical depth. Typical numbers
are $10^8K$ for temperature and 0.01 for optical depth.
Positive detections in the RJ-tail 
have by now been achieved for more than 30 clusters with central 
decrements exceeding $500 \mu K$ (Carlstrom et al. 1999). This is a huge signal that can 
easily swamp the lensing signal. 
Convolving with the optimal filter reduces the level of fluctuations 
and even eliminates them for axi-symmetric profiles.
However most of the clusters are  not axi-symmetric and 
for reasonable ellipticities the remaining contamination  could still be above
the expected signal in the center.
One may further reduce this contamination by eliminating the central 
region of the cluster in the analysis. Most of the SZ signal is coming 
from the inner 1-2' radius, while the lensing signal extends well beyond that.
To model the importance of the inner part of the cluster we repeated
the noise analysis excluding the lensing information from the inner 4'
region. This increased the variance by 40\% and so does not 
significantly reduce the sensitivity, while reducing the level of SZ
signal by a factor of a few.

Further reduction of this contamination is achieved
by observing at 217 GHz.
Although this frequency is a zero crossing for thermal SZ
in the non relativistic limit, for most
of the clusters with large signal relativistic effects are not 
negligible. This causes 
the zero crossing to scale linearly with the gas temperature (Rephaeli 1995).
If the cluster is isothermal and its temperature can 
be measured from X-ray measurements then one can correct for this effect.
If the cluster is not isothermal as suggested by recent ASCA measurements 
(Markevitch et al. 1996)
then this will induce further fluctuations in the map which can be at a
10$\mu K$ level. These fluctuations can be reduced using lensing
filter combined with exclusion of the center and so do not appear to be 
a major source of confusion.

\subsection{Kinetic SZ effect from the cluster}

Even if SZ effect from hot electrons vanishes at 217 GHz there is
another imprint of the CMB photons scattering off cluster electrons, 
caused by electron bulk motion. This is caused by the Doppler effect
and its magnitude is given by the product of optical depth, typically 
around $\tau=0.01$ in the center and radial velocity of electrons in 
the cluster. The latter is dominated by the bulk motion of cluster $v_r$
with a typical value of $v_r=300km/s$. This gives 
the typical magnitude of the effect in the center around 30 $\mu K$, 
somewhat larger than
the lensing signal. Note that
the two have the same frequency dependence and cannot be distinguished 
using this information.
However, just like in the case of thermal SZ 
kinetic SZ is much more centrally concentrated than the lensing effect. 
The temperature profile in the presence of both 
effects is shown in figure \ref{fig1}d for the case of the 
NFW profile (we are assuming that gas traces dark matter 
outside the cluster core). 
At larger separations from the center kinetic SZ becomes negligible, 
while lensing effect remains strong, so the two effects can be 
separated. Excluding the central portion of the cluster and using the  
optimal filter we find confusion levels of a few $\mu K$.

Another potential source of contamination are the bulk motions
within the cluster. If the cluster is not relaxed due to a recent 
merger this can produce significant internal motions of the gas
(of the order of 500 km/s, Haehnelt and Tegmark 1996).
The corresponding Doppler effect on the CMB can act as an 
additional source of fluctuations.  
Filtering reduces
the noise level when the large scale CMB gradient is known, with the
residual contamination at the $\mu K$ level.

\subsection{SZ and OV along the line of sight}

In addition to the thermal and kinetic SZ effect from the cluster itself
there is also the contribution from other objects along the line of 
sight. Kinetic SZ is sometimes divided into
a contribution from quasi-linear structures
called Ostriker-Vishniac (OV) effect and a contribution from nonlinear 
structures (kinetic SZ). 
All these will be a source of noise uncorrelated with the cluster 
itself. The magnitude of these contributions is somewhat model dependent. 
Both of them can be at a level of few $\mu K$ on arcminute scales, 
with thermal SZ being typically a few times stronger.
However, since thermal SZ vanishes at 217 GHz 
(relativistic corrections are likely to be 
negligible for smaller halos contributing to the line of sight SZ), 
OV and kinetic SZ may be more important as a source of 
confusion at this frequency. 
To estimate their effect we have used the power spectra of 
thermal SZ as given in Persi et al. (1995) and OV given in Hu (1999). 
Thermal SZ power spectrum grows roughly proportional 
to $l$ and exceeds CMB around $l \sim 2000$ at low frequencies. 
Kinetic SZ is somewhat lower, but also grows at high $l$ in a similar fashion. 
Only on intermediate scales do the two exceed the combined CMB and instrument 
noise power spectrum.
Their individual contribution to 
the lens filter variance varies as a function of angular scale. 
For the 5' beam with 12 $\mu K$ noise per pixel 
the contribution from thermal SZ can double the 
variance from CMB and noise making the total 20 $\mu K$. At this angular resolution it 
is necessary to work at 217 GHz frequencies to reduce thermal SZ 
contamination, although kinetic SZ/OV still increases the variance somewhat.
For a 1' beam with 5 $\mu K$ noise per pixel 
the contributions from SZ and OV are lower and do not significantly 
change the rms noise. This is because instrument noise dominates the confusion. 
Only with a more sensitive detector would these contributions become important on 
these scales. 

\subsection{Dust emission}
Dust emission can arise from three separate sources. First there is 
the emission from our own galaxy. This contribution is fairly smooth, scaling as
$C_l \propto l^{-3}$,
and does not add significant power on small scales.
A reasonable 
estimate for noise variance 
is 10$\mu K$ at $l=10$ for 217 GHz channel (Tegmark et al. 1999), dropping 
significantly at lower frequencies. Even at 217 GHz its power spectrum is 
below the CMB power spectrum everywhere except at very low $l$. This 
foreground is therefore less problematic than the primary CMB and its 
inclusion does not change significantly the conclusions above. As a 
caution we should note that this conclusion is based only on the power 
spectrum
analysis, while dust emission can be strongly non Gaussian. There are 
regions where dust emission can be significantly larger than the above
analysis would suggest. An example discussed below is in the field 
of A2163.

Another source of dust emission are the infrared 
sources along the line of sight. These have been modeled by Toffolatti et al.
(1998) and Guiderdoni et al. (1999). 
The overall contribution from the point sources
to the power spectrum depends on the flux limit
of the resolved sources.
Strong point-like sources can 
be removed from the data as outliers, leaving  
the fluctuations 
produced by the unresolved sources. These Poisson fluctuations 
give a white noise power spectrum with rms fluctuations 
of the order of the flux limit converted to $\mu K$ in a beam
area times square root of number of removed sources per beam area.
Adopting conservative modeling as in Tegmark et al. (1999) we 
find that at 217 GHz the rms variance including point sources 
can reach 25 $\mu K$ for 1' beam. This is reduced somewhat at 
larger angular scales, but there thermal SZ and CMB combined
prevent one from reducing the noise below 10 $\mu K$. These 
results indicate that more sophisticated modeling of point 
sources will be necessary to reduce their contribution to 
acceptable levels. This can be achieved by using either higher 
frequencies 
or higher angular resolution to identify these sources.

Finally there is also the possibility of the 
dust emission from the cluster itself. Such emission
could explain recent recent sub-mm observation of A2163 by Pronaos
(Lamarre et al. 1998) and may extend into the 200GHz range, although 
alternative explanation with galactic dust is just as likely. 
This would complicate the 
assertion that the 217 GHz frequency is the optimal one for identifying 
this effect. If the cluster dust emission at this frequency is still strong it 
may exceed the lensing signal at least in the center. 
It seems however unlikely
that a strong dust component would also be present at larger separations
from the center. 

\subsection{Radio point sources}

At low frequencies the main source of confusion are radio point 
sources (we ignore here free-free and synchrotron which typically 
do not exceed the CMB power spectrum and so are subdominant in 
contamination). Their modeling has also been 
presented in Toffolatti et al. (1998). At 30 GHz their contribution 
to the power spectrum using only internal identification (based
on their identification as outliers in the flux distribution) is 
about 100 times larger than the point source contribution from IR 
sources at 217 GHz. The variance on the filtered profile is around 
100 $\mu K$ for 5' beam and twice that for 1' beam. This is of course
well known to observers operating in this frequency range, who 
routinely employ higher sensitivity multi frequency 
observations in the same region 
to eliminate point source contamination. This can reduce 
the variance from point sources 
below the instrument noise, 15-40 $\mu K$ for currently most 
sensitive experiments (Carlstrom et al. 1999).
It remains to be seen however whether using this additional information 
can reduce the contamination to the required level of a few $\mu K$.

\subsection{Gravitational lensing}

Distortion of the background CMB is caused by all the matter distribution 
along the line of sight, so there will be additional fluctuations added 
on top of the effect from the cluster. The effect can be partially modeled
by using the lensed instead of unlensed power spectrum in the estimate 
of confusion from primary anisotropies discussed above. This 
gives additional noise contribution at the level below 1 $\mu K$ and 
so would appear not to significantly contaminate the signal. 
This approach however
underestimates this contribution, because the generated CMB power 
is also correlated with the CMB gradient and the lens filter does not 
eliminate it as efficiently as the power spectrum analysis would indicate. 
As shown in Zaldarriaga and Seljak (1999b)
lensing along the line of sight primarily generates power on 
scales smaller than the cluster, with rms amplitude of a few 
$\mu K$. Averaging over the expected cluster profile reduces 
this noise to negligible levels compared to other sources.

\subsection{Atmosphere noise}
In addition to the sources of noise described above for ground 
experiments there is also the
atmosphere noise arising from atmosphere temperature (around 15K)
and atmosphere fluctuations. The first can be modeled as a white noise
and has similar properties to the instrument noise. For bolometer 
arrays it dominates the noise at small scales, so sufficient long 
integrations are needed to reduce it to acceptable levels. Atmospheric
fluctuations are more correlated and their precise estimate depends 
on the specifics of the detector, site, weather etc. They can be 
reduced significantly using interferometric techniques and it is expected
that they can be reduced
to a few $\mu K$ levels required. 

\section{Conclusions}

The characteristic signature of a cluster gravitationally lensing
a smooth CMB gradient
allows one to search and identify this 
effect among many other possible sources of fluctuations at small scales. 
The signal is small, at the level around 10 $\mu K$ in the center 
of the most promising clusters
and a fraction of that away from it.  Other sources of anisotropies 
almost certainly exceed this  
signal in the absence of filtering. Noise filtering over the expected profile 
can reduce the 
noise contamination to acceptable levels if the initial guess about the 
cluster profile is sufficiently close to the real one. The 
final signal to noise depends sensitively on the amplitude of various 
sources of fluctuations including instrument noise, all of which are 
still rather uncertain at present. The most promising among the existing
or planned experiments are 
small scale interferometers with arcminute resolution and a few $\mu K$
noise per pixel sensitivity operating at frequencies close to 217 GHz.
Planck experiment with 5' beam at 217 GHz could also provide  
detection in some cases. 

Despite the signal being weak, this method of detecting cluster 
signature has some advantages over the current methods. First,
like for the
SZ effect the strength does not drop significantly with redshift, 
except for the slow decrease from the $D_{LS}/D_{OS}$ ratio. 
In addition, the signal depends on the deflection 
angle, which is more sensitive to the 
outer parts of the cluster as those probes that depend on the 
projected density. For the NFW profile the signal drops by about a 
factor of 2 between $\theta_s$ and $10\theta_s$ (which is beyond 
the virial radius), while the projected surface density drops by 
almost two orders of magnitude (see figure \ref{fig2}). 
Given that this is one of the few probes sensitive to the outer 
parts of the cluster 
it seems worth pursuing it
with the next generation of CMB experiments.

\smallskip
We thank L. Page for useful discussions.  
M.Z. is supported by NASA through Hubble Fellowship grant
HF-01116.01-98A from STScI,
operated by AURA, Inc. under NASA contract NAS5-26555.
U.S. is supported by NASA grant NAG5-8084.

\end{document}